\documentclass[twocolumn,12pt]{article}

\usepackage{amssymb,amsmath,graphicx}

\newcommand{\tab}{\hspace{5mm}}

\newcommand{\vect}[1]{\textbf{#1}}

\begin{document}

\twocolumn[    \begin{@twocolumnfalse}
\begin{center}
\textbf{{\Large Fourvector Algebra}}\\
\vspace{0.5 cm}

\textbf{\textit{Diego Sa\'{a}} $^{1}$}\\
\vspace{0.5 cm}
{(1)~Escuela Polit\'{e}cnica Nacional. Quito --~Ecuador.\\ e-mail: dsaa@server.epn.edu.ec}\\

\end{center}

\begin{abstract}
{The algebra of fourvectors is described. The fourvectors are more appropriate than the Hamilton quaternions for its use in Physics and the sciences in general. The fourvectors embrace the 3D vectors in a natural form. It is shown the excellent ability to perform rotations with the use of fourvectors, as well as their use in relativity for producing Lorentz boosts, which are understood as simple rotations.}
\end{abstract}

\textit{PACS}: 02.10.Vr\\

\textbf{Key words}: fourvectors, division algebra, 3D-rotations, 4D-rotations\\

\vspace{0.5 cm}

\pagebreak

\vspace{1.0 cm}

\end{@twocolumnfalse}  ]

\newpage

\tableofcontents

\newpage

\section{Introduction }

\subsection{General}

In this paper it is suggested the use of fourvectors with the purpose of replacing the 3D vectors and the quaternions. Because the fourvectors contain the three dimensional vectors and can be considered a formalization of them.\\
The discovery of the quaternions is attributed to the Irish mathematician William Rowan Hamilton in 1843 and they have been used for the study of several areas of Physics, such as mechanics, electromagnetism, rotations and relativity \cite{silberstein1912}, \cite{horn02}, \cite{deleo96}, \cite{baez01}, \cite{mukundan02}, \cite{greiter05}. James Clerk Maxwell used the quaternion calculus in his Treatise on Electricity and Magnetism, published in 1873 \cite{maxwell54}. An extensive bibliography of more than one thousand references about \textit{Quaternions in mathematical physics} has been compiled by Gsponer and Hurni \cite{gsponer06}. 

The modern vectors were discovered by the Americans Gibbs and Heaviside between 1888 and 1894. Their work may be considered a sort of combination of quaternions and ideas developed around 1840 by the German Hermann Grassman. The notation was primarily borrowed from quaternions but the geometric interpretation was borrowed from Grassman's system. 

By the end of the nineteenth century the mathematicians and physicists were having difficulty in applying the quaternions to Physics.

Ryan J. Wisnesky  \cite{wisnesky04} explains that ``The difficulty was a purely pragmatic one, which Heaviside was expressing when he wrote that \textit{there is much more thinking to be done} [\textit{to set up quaternion equations}]. In principle, most everything done with the new system of vectors could be done with quaternions, but the operations required to make the quaternions behave like vectors added difficulty to using them and provided little benefit to the physicist.''

``Gibbs was acutely aware that quaternionic methods contained the most important pieces of his vector methods.'' \cite{wisnesky04}

After a heated debate, ``by 1894 the debate had largely been settled in favor of modern vectors'' \cite{wisnesky04}. 

Alexander MacFarlane was one of the debaters and seems to have been one of the few in realizing what the real problem with the quaternions was. ``MacFarlane's attitude was intermediate - between the position of the defenders of the Gibbs\textendash Heaviside system and that of the quaternionists. He supported the use of the complete quaternionic product of two vectors, but he accepted that \textit{the scalar part of this product should have a positive sign}.
According to MacFarlane the equation $\vect{j k}=\vect{i}$ was a convention that should be interpreted in a geometrical way, but \textit{he did not accept that it implied the negative sign of the scalar product}''. \cite{silva02} (The emphases are mine).\\
He incorrectly attributed the problem to a secondary and superficial matter of representation of symbols, instead of blaming to the more profound definition of the quaternion product.
``MacFarlane credited the controversy concerning the sign of the scalar product to the conceptual mixture done by Hamilton and Tait. He made clear that \textit{the negative sign came from the use of the same symbol to represent both a quadrantal versor and a unitary vector}. His view was that different symbols should be used to represent those different entities.'' \cite{silva02} (The emphasis is mine).\\

At the beginning of the twenty century, Physics in general, and relativity theory in particular, was lacking an appropriate mathematical formalism to represent the new physical quantities that were being discovered. But, despite the fact that all physical variables such as space-time points, velocities, potentials, currents, etc., were recognized that must be represented with four values, the quaternions were not being used to represent and manipulate them. It was necessary to develop some new mathematical devices to manipulate such variables. Besides vectors, other systems such as tensors, spinors, matrices and geometric algebra were developed or used to handle the physical variables.\\

During the twenty century we have witnessed further efforts to overcome the difficulties remaining, with the development of other algebras, which recast several of the ideas of Grassman, Hamilton and Clifford in a slightly different framework. An example in this direction is Hestenes' Geometric Algebra in three dimensions and Space Time Algebra in four dimensions. \cite{hestenes66}, \cite{hestenes84}, \cite{lasenby96}, \cite{doran03} \cite{hestenes02}

The commutativity of the product was abandoned in all the previous quaternions and in some algebras, such as the one of Clifford. According to Gaston Casanova \cite{casanova02} ``It was the English Clifford who carried out the decisive path of abandoning all the commutativity for the vectors but conserving their associativity.'' ``This algebra absorbs the Hamilton quaternions, the Girard's complex quaternions, the cross product and the complex numbers, the hyperbolic numbers and the dual numbers.'' \cite{casanova02}. Also the Hestenes' ``geometric product'' conserves associativity \cite{hestenes02}. In this sense, the associativity of the product is finally abandoned in the fourvector algebra proposed in the present paper. This means that the fourvectors do not constitute a Clifford Algebra \cite{baez01} or a Geometric Algebra \cite{aragon1997}. This is a collateral effect of the proposed algebra, and constitutes a hint about the form the fourvectors should handle, for example, a sequence of rotations; besides, the complex numbers are not handled as in the Hamilton quaternions, where the real number is put in the scalar part and the imaginary in the vector part, but a whole complex number can be put in each component, so it is possible to have up to four complex numbers in each fourvector. But, what is more important, it is known that in quantum mechanics, observables do not form an associative algebra, so this could be the natural algebra for Physics.

The proposed algebra could have been already developed, around 1900, under the name of \textit{hyperbolic quaternion}, which is a mathematical concept introduced by Alexander MacFarlane of Chatham, Ontario. The idea was dismissed for its failure to conform to associativity of multiplication, but it has a legacy in Minkowski space and as an extension of ``split-complex numbers''. Like the quaternions, it is a vector space over the real numbers of dimension 4. There is only the mention to such quaternions but no accessible references to confirm if those quaternions satisfy the same algebraic rules given in the following for the fourvectors. However our intent is to convince the reader that the presented here is one of the most important mathematical tools for Physics.\\

The fourvector representation is without a doubt a more unified theory in comparison to the classical vector or matrix representations.\\

The use of fourvectors allows discerning constants, variables and relations, previously unknown to Physics, which are needed to complete and make coherent the theory.\\

The vectors have lost some ground in favor of the Hamilton quaternions due to the lack of an appropriate 4D-algebra. For example, Douglas Sweetser, who has worked extensively in the application of Hamilton quaternions to many possible physical areas, in general with very little success, sustains these opinions: ``Today, quaternions are of interest to historians of mathematics. Vector analysis performs the daily mathematical routine that could also be done with quaternions. I personally think that there may be 4D roads in physics that can be efficiently traveled only by quaternions.'' \cite{sweetser}\\

In fact those 4D roads should be traveled only by properly handled fourvectors. It has been an old dream to express the laws of Physics using quaternions. But this attempt has been plagued with recurring pitfalls for reasons until now unknown to both physicists and mathematicians. The quaternions have not been making problem solving easier or simplifying the equations. Very often the Hamilton quaternions require an extreme hability to guess when and where a quaternion needs to be conjugated, in order to obtain some particular result.

I believe that this has been due to an internal problem in the mathematical structure of the Hamilton quaternions, which I will try to reveal in the present paper. The correction of such problem constitutes a new non-commutative, non-associative normed algebraic structure with which it is possible to work with fourvectors in an improved way relative to the Hamilton, Pauli or Dirac quaternions, geometric algebra, space\textendash time algebra and other formalisms. \\

In the present paper, in particular, the present author exhibits the application of the fourvectors to 3D and 4D rotations, which requires a reformulation of the Hamiltonian mathematics.\\

The powerful \textit{Mathematica}\textregistered \, package includes a standard algebra package for the manipulation of the Hamilton quaternions. I have borrowed from that package the symbol, as double asterisk, to represent the fourvector product. It is easy to modify the cited package to handle the fourvectors, as well as to permit not only their numeric but also symbolic and complex manipulation. Though I have still not been able to figure out a simple way to reproduce the trigonometric and exponential functions for the fourvectors (if at all possible), which that package allows to compute for the Hamilton quaternions.\\

In the following three subsections, a cursory revision is made of the Hamilton, Pauli and Dirac quaternions, for an easy comparison with fourvectors. In section 2 the fourvectors are presented. Section 3 can be skimmed by the mathematician, since it is mostly classic algebra. Finally, in section 4 the formulae needed to perform rotations and reflections with fourvectors is presented.

\subsection{The Hamilton quaternions}

Quaternions are ``four-dimensional numbers'' of the 
form \cite{waser01}:

\begin{align}
\vect{A} &= a + \vect{i}\, a_{x} + \vect{j}\, a_{y} + \vect{k}\, a_{z},\\
\vect{B} &= b + \vect{i}\, b_{x} + \vect{j}\, 
b_{y} + \vect{k}\, b_{z}
\nonumber
\end{align}

\noindent
where the basis elements 1, \textbf{i}, \textbf{j}, \textbf{k} satisfy the relations:
\begin{equation}
\label{hbasis1}
\textbf{i}^{2} = \textbf{j}^{2} = \textbf{k}^{2} = \vect{i} \vect{j}\, \vect{k} = -1
\end{equation}

and also:
\begin{align}
\label{hbasis2}
\vect{i} \,\vect{j} &= - \vect{j} \, \vect{i} = \vect{k},\\
\vect{j} \,\vect{k} &= - \vect{k} \,\vect{j} = \vect{i},\\
\label{hbasis4}
\vect{k} \,\vect{i} &= - \vect{i} \,\vect{k} = \vect{j}.
\end{align}

Here 1 is the usual real unit; its product with \textbf{i}, \textbf{j} or \textbf{k} 
leaves them unchanged. \\
Thus, since the products of the basis elements are non-commutative, 
we have in general\\ $\vect{A}$**$\vect{B} \neq \vect{B}$**$\vect{A}$, where the double asterisk represents 
quaternion multiplication. Under these conditions, quaternion 
multiplication is associative, so that (\vect{A}**\vect{B})**\vect{C} = \vect{A}**(\vect{B}**\vect{C}) 
for any three quaternions \vect{A}, \vect{B}, \vect{C}.

The sum of two quaternions is

\begin{equation}
\begin{split}
\label{hsum}
\vect{A} + \vect{B} =& (a + b) + \vect{i} (a_{x} + b_{x}) + \\
 &\vect{j} (a_{y} + b_{y}) + \vect{k} 
(a_{z} + b_{z}),
\end{split}
\end{equation}

\noindent
and, using relations \eqref{hbasis1} and \eqref{hbasis2}-\eqref{hbasis4}, the product is given by:

\begin{equation}
\begin{split}
\label{hproduct}
\vect{A}**\,\vect{B} =& \; (a b - a_{x} b_{x} - a_{y} b_{y} - a_{z} b_{z}) +\\
& \vect{i}\, (a b_{x} + a_{x} b + a_{y} b_{z} - a_{z} b_{y}) +\\
& \vect{j}\, (a b_{y} - a_{x} b_{z} + a_{y} b + a_{z} b_{x}) +\\
& \vect{k}\, (a b_{z} + a_{x} b_{y} - a_{y} b_{x} + a_{z} b).
\end{split}
\end{equation}

The notation of three-dimensional vector analysis furnish a useful 
shorthand for quaternion operations. Regarding \vect{i}, \vect{j}, \vect{k} 
as unit vectors in a Cartesian coordinate system, we interpret the quaternion \vect{A} as comprising the scalar part \textit{a} and the vector part \\ \vect{a} = \vect{i} a$_{x}$ + \vect{j} a$_{y}$ + \vect{k} a$_{z}$. Then we write in the simplified form \vect{A} = (\textit{a}, \vect{a}). With this notation, the sum \eqref{hsum} and the product \eqref{hproduct} may more compactly be expressed as:
\begin{align}
\vect{A} + \vect{B} = &( \textit{a} + \textit{b}, \vect{a} + \vect{b} )
\end{align}
\begin{align}
\begin{split}
\vect{A}**\,\vect{B} =&( \textit{a b} - \vect{a} \cdot \vect{b}, \textit{a} \, \vect{b} + \textit{b} \, \vect{a} + \vect{a} \times \vect{b})
\end{split}
\end{align}

where the usual rules for vector sum and dot and cross products 
are being invoked.

According to the mathematicians, the Hamilton quaternions are mathematical structures which combine properties of complex numbers and vectors.

\subsection{The Pauli quaternions}

The Hamilton multiplication rules differ from the Pauli matrices 
rules only by the explicit appearance of the fourth basis element.\\

The basis elements of the Pauli quaternion space are denoted by $\vect{s}_{\mathbf{1}}, \vect{s}_{\mathbf{2}}, \vect{s}_{\mathbf{3}}, \vect{s}_{\mathbf{4}}.$\\

They obey the following multiplication rules, comparable to \eqref{hbasis1}-\eqref{hbasis4}:
\begin{equation}
\label{paulibasis}
\begin{split}
&\vect{s}_{\mathbf{1}}^{2} =\vect{s}_{\mathbf{2}}^{2} =\vect{s}_{\mathbf{3}}^{2} = -
\vect{s}_{\mathbf{4}}^{2} =\ -\textbf{1}\\
&\vect{s}_{\mathbf{1}} \, \vect{s}_{\mathbf{2}} =-\textbf{s}_{\mathbf{2}} \, \vect{s}_{\mathbf{1}} =\textbf{s}_{\mathbf{3}}\\
&\vect{s}_{\mathbf{3}} \,
\vect{s}_{\mathbf{1}} =-\vect{s}_{\mathbf{1}} \,  \vect{s}_{\mathbf{3}} =\textbf{s}_{\mathbf{2}}\\
&\vect{s}_{\mathbf{2}} \, \vect{s}_{\mathbf{3}} =- \vect{s}_{\mathbf{3}} \, \vect{s}_{\mathbf{2}} =\textbf{s}_{\mathbf{1}}\\
&\vect{s}_{\mathbf{4}} \, \vect{s}_{\mathbf{k}} =\; \; \vect{s}_{\mathbf{k}} \, \vect{s}_{\mathbf{4}} \,=\vect{s}_{\mathbf{k}} 
, \quad (k= 1, 2, 3, 4)
\end{split}
\end{equation}

These rules are satisfied, in particular, by the Pauli spin matrices (only the first three bear this name, because $\sigma _{4}$ serves to form the identity matrix) \cite{hestenes02}, \cite{horn02}, \cite{beil03} \cite{gibbon07}:

\begin{equation}
\begin{split}
\sigma _{1} = 
\begin{bmatrix}
0 & \;\; 1 \\
1 & \;\; 0
\end{bmatrix},\;
\sigma _{2} =
\begin{bmatrix}
0 & -\textit{i} \\
\textit{i} &  \;\;0
\end{bmatrix} \\
\sigma _{3} = 
\begin{bmatrix}
1 & 0 \\
0 & -1
\end{bmatrix},\; 
\sigma _{4} =
\begin{bmatrix}
\textit{i} & \;\;0 \\
0 & \;\;\textit{i}
\end{bmatrix}
\end{split}
\end{equation}

\noindent
where ``\textbf{1}'' in \eqref{paulibasis} represents the identity matrix, \textit{i} the imaginary unit and $\vect{s}_{i} = \,-\,\textit{i}\; \sigma_{i}$, for $\textit{i} \in {\{1, 2, 3, 4\}}$.\\

The Pauli quaternions evidence one difference with respect to the classical Hamilton quaternions, being the need of matrices, which in some cases have imaginary units \textit{i}. \\

The sum of two Pauli quaternions is of the same form as the given for the Hamilton quaternions and its product, using \eqref{paulibasis}, becomes:

\begin{equation}
\begin{split}
\vect{A}**\,\vect{B} = &\vect{s}_{\mathbf{4}}(a b - a_{x} b_{x} - a_{y} b_{y} - a_{z} b_{z}) +\\
&\vect{s}_{\mathbf{1}} (a b_{x} + a_{x} b + a_{y} b_{z} - a_{z} b_{y}) +\\
&\vect{s}_{\mathbf{2}} (a b_{y} - a_{x} b_{z} + a_{y} b + a_{z} b_{x}) +\\
&\vect{s}_{\mathbf{3}} (a b_{z} + a_{x} b_{y} - a_{y} b_{x} + a_{z} b).
\end{split}
\end{equation}

In a compact form, the product for the Pauli quaternions has exactly the same form as the Hamilton quaternions and, therefore, have the same problems as these:

\begin{equation}
\vect{A}**\,\vect{B} = ( a \, b\,  - \; \textbf{a} \cdot \textbf{b}, \, a \, \textbf{b}\,  +\,  b \, \textbf{a} +\,  \textbf{a} \times \textbf{b})
\end{equation}

\subsection{Dirac matrices}
The Dirac matrices must satisfy the Klein-Gordon equation, the following relations should be satisfied by the Dirac matrices \cite{yamamoto05}:

\begin{equation}
\begin{split}
\label{kleingordon}
&\alpha _{i} \alpha _{j} +\alpha _{j} \alpha _{i} =2\delta _{ij},\\ 
&\alpha _{i} \beta +\beta \alpha _{i} =0,  \quad(\textit{i} =1, 2, 3)\\
&\alpha _{i} ^{2} =\beta ^{2} =I 
\end{split}
\end{equation}

where \textit{I} represents a $N\times N$ unit matrix.\\

For $2\times 2$ matrices, only three anti-commuting matrices exist (the Pauli matrices). Thus the smallest dimension allowed for the Dirac matrices is $N=4$. If one matrix is diagonal, the others can not be diagonal or they would commute with the diagonal matrix. We can write a representation that is hermitian (a matrix is hermitian if it is equal to the conjugate of its transpose), traceless (trace equal zero), and has eigenvalues of $\pm 1$:

\begin{equation}
\alpha_{i} =
\begin{pmatrix}
0 & \sigma_{i}  \\
\sigma_{i} & 0
\end{pmatrix}
 , \quad (\textit{i} =1, 2, 3)
\end{equation}
and
\begin{equation}
\beta =
\begin{pmatrix}
I & \; 0 \\
0 & -I
\end{pmatrix}
\end{equation}

where $\sigma _{i} $ are the $2\times 2$ Pauli matrices and \textit{I} is the $2\times 2$ unit matrix.\\

Finally, we are ready to define the Dirac's gamma matrices out of $\alpha_{i}$ and $\beta$:
\begin{equation}
\gamma^{0}\equiv \beta,\quad \gamma^{i}\equiv \beta\,\alpha_{i}\; (i=1,2,3)
\end{equation}

These matrices satisfy the relations:
\begin{equation}
\label{dirac}
(\gamma^{0})^{2} = 1, \quad(\gamma^{i})^{2} = -1,
\end{equation}
\noindent
and all four matrices anticommute among themselves. These relations are comparable to the Hamilton basis \eqref{hbasis1}-\eqref{hbasis4} and Pauli basis \eqref{paulibasis}, except for the exchange of secondary importance in the signature of the gamma matrices, $(+,-,-,-)$, to the signature satisfied by the Hamilton and Pauli bases: $(-,+,+,+)$.

\section{The fourvectors}

The fourvectors are four-dimensional numbers of the form

\begin{equation}
\vect{A} = \vect{e} \, a_{t} + \vect{i} \, a_{x} + \vect{j} \, a_{y} + \vect{k} \, a_{z}
\end{equation}

or, assuming that the order of the basis elements is the indicated, those basis elements can be suppressed and included implicitly in a notation similar to a vector or 4D point:
\begin{equation}
\vect{A} = (a_{t}, a_{x}, a_{y}, a_{z})
\end{equation}

Where the elements of the fourvector can be any integer, real, imaginary or complex numbers.\\

The four basis elements \vect{e}, \vect{i}, \vect{j}, \vect{k} satisfy the relations:
\begin{equation}
\label{rel1}
\vect{e}^{2} = \vect{i}^{2} = \vect{j}^{2} = \vect{k}^{2} = \vect{e} = \vect{e} \, \vect{i} \, \vect{j} \, \vect{k}
\end{equation}
\noindent
The following rules are satisfied by the basis elements:

\begin{equation}
\begin{split}
\label{rel2}
&\vect{e i} = - \vect{i e} = \vect{i},\\
&\vect{e j} = - \vect{j e} = \vect{j},\\
&\vect{e k} = - \vect{k e} = \vect{k},\\
&\vect{i j}\, = - \vect{j\, i} = \vect{k},\\
&\vect{j k} = - \vect{k j} = \vect{i},\\
&\vect{k i} = - \vect{i k} = \vect{j}.
\end{split}
\end{equation}
 
The group of relations \eqref{rel2} gives an important operational mechanism to reduce any combination of two or more indices to at most one.

The ``\vect{e\,i\,j\,k}'' bases characterize the fourvector 
product as not commutative but, what is more important and different with respect to the previous Hamilton and Pauli quaternions as well as to the Clifford Algebra (see \cite{casanova02}, p. 5 axiom 3), the product is not always associative. For example consider the product of the following four symbols ``$\vect{i\,e\,j\,k}$'', in the order given; if, with the use of \eqref{rel2}, we first reduce ``$\vect{i e}$'' to $-\vect{i}$ then ``$-\vect{i j}$'' to $-\vect{k}$ and finally ``$-\vect{k k}$'' to $-\vect{e}$ we obtain one result; but, if we first reduce the two middle basis elements ``$\vect{e j}$'' to $\vect{j}$, then ``$\vect{i j}$'' to $\vect{k}$ and then ``$\vect{k k}$'' to $\vect{e}$, we get the same result but with the sign changed. \\

If we put these rules into a multiplication table they look in the following way:

\begin{tabular}{|l|l|l|l|l|}
\hline
\textbf{**} & 
\; \textbf{e} & 
\; \textbf{i} & 
\; \textbf{j} & 
\; \textbf{k} \\
\hline
\; \textbf{e} & 
\; {e} & 
\; {i} & 
\; {j} & 
\; {k} \\
\hline
\; \textbf{i} & 
--{i} & 
\; {e} & 
\; {k} & 
--{j} \\
\hline
\; \textbf{j} & 
--{j} & 
--{k} & 
\; {e} & 
\; {i} \\
\hline
\; \textbf{k} & 
--{k} & 
\; {j} & 
--{i} & 
\; {e} \\
\hline
\end{tabular}\\

Let us assume that each basis unit is affixed its proper number (as components of a tensor):\\
 $(\vect{e}, \vect{i}, \vect{j}, \vect{k}) \rightarrow 
(q_{0}, q_{1}, q_{2}, q_{3}) = q,$ \\
then the fourvector multiplication satisfies the following compact relation, where the symbol $\chi$ has some similitude with the Levi-Civita symbol. Refer to \cite{sinclair05} for the meaning of such symbol.\\

$q_{i} q_{k} =\delta _{i k} q_{0} +\chi _{i k }^{j} q_{j} 
 \tab  i, j, k$ = 0, 1, 2, 3\\

\subsection{Discussion}

Tait and Gibbs discussed about the relative merits of the ``vector product'' against the ``quaternion product''. Tait appeared to gain a slight advantage by pointing out that quaternion products are associative, whereas the cross product is not. Tait used the following example to reveal the supposed deficiency of the vectors \cite{wisnesky04}:
\begin{equation}
\vect{i}\times (\vect{j}\times \vect{j})=\vect{0}\neq (\vect{i}\times \vect{j})\times \vect{j}=-\vect{i}
\end{equation}

Note that the four-vector product proposed in the previous section, although not associative, or precisely because of that, gives the correct result for the problem at hand:
\begin{equation}
\begin{split}
\vect{i} \times (\vect{j} \times \vect{j}) \rightarrow \vect{i} \times \vect{e} \rightarrow -\vect{i}\\
(\vect{i} \times \vect{j}) \times \vect{j} \rightarrow \vect{k} \times \vect{j} \rightarrow -\vect{i}
\end{split}
\end{equation}

The fourvectors have extensive applications in electrodynamics and relativity. The present author believes that the use of the fourvectors, with the proposed algebra, can replace advantageously the matrices, vectors and tensors in representation. Some of the advantages proposed for the Hamilton quaternions, Geometric Algebra and Space-Time Algebra, which are also extended to the fourvectors, are:
\begin{enumerate}
\item Fourvectors can express rotation as a rotation angle about a rotation axis. This is a more natural way to perceive rotation than Euler angles \cite{dam98}.
\item Non singular representation (compared with Euler angles, for example)
\item More compact (and faster) than matrices. For computation with rotations, fourvectors offer the advantage of requiring only 4 numbers of storage, compared with 9 numbers for orthogonal matrices \cite{salamin1979}. Composition of rotations requires 16 multiplications and 12 additions in fourvector representation, but 27 multiplications and 18 additions in matrix representation...The fourvector representation is more immune to accumulated computational error. \cite{salamin1979}.
\item Every fourvector formula is a proposition in spherical (sometimes degrading to plane) trigonometry, and has the full advantage of the symmetry of the method \cite{tait1886}.
\item Unit fourvectors can represent a rotation in 4D space.
\item Fourvectors are important because of their ``all-attitude'' capability and numerical advantages in simulation and control \cite{stevens03}.
\end{enumerate}

Quaternions have been often used in computer graphics (and associated geometric analysis) to represent rotations and orientations of objects in 3D space. This chores should be now undertaken by the fourvectors, which are more natural, and more compact than other representations such as matrices, and operations on them such as composition can be computed more efficiently. Fourvectors, as the previous quaternions, will see uses in control theory, signal processing, attitude control, physics, and orbital mechanics, mainly for representing rotations/orientations in three dimensions. The spacecraft attitude-control systems should be commanded in terms of fourvectors, which should also used to telemeter their current attitude. The rationale is that combining many fourvectors transformations is more numerically stable than combining many matrix transformations.

\subsection{Complex fourvectors}

The only difference with respect to the ordinary fourvectors is that the elements are not purely real but complex numbers.\\
The collection of all complex fourvectors forms a vector space of four complex dimensions or eight real dimensions. Combined with the operations of addition and multiplication, this collection forms a non-commutative and non-associative algebra. There is no difficulty in obtaining the multiplicative inverse of a complex fourvector, when it exists, within the fourvector algebra suggested below. However, there are complex fourvectors different from zero whose norm is zero. Therefore the complex fourvectors do not constitute a division algebra.\\

It is important to realize that the relations needed by the Klein-Gordon equation \eqref{kleingordon}, are directly satisfied by the purely real fourvectors, whereas the relations needed by the Dirac equation \eqref{dirac}, are satisfied by the fourvectors constituted of imaginary components in the vector part.\\

This seems to suggest that there are two different kinds of physical entities, although closely related, which need respectively the real and the imaginary representations. This insight appears potentially useful for Physics.\\

\section{Fourvector algebra}

The \textit{sum} of two fourvectors is another fourvector where each component has the sum of the corresponding argument components:

\begin{equation}
\begin{split}
\vect{A} + \vect{B} = &\vect{e} (a_{t} + b_{t}) + \vect{i} (a_{x} + b_{x})+\\
&\vect{j} (a_{y} + b_{y}) + \vect{k} (a_{z} + b_{z})
\end{split}
\end{equation}

The \textit{difference} of two fourvectors is defined similarly:
\begin{equation}
\begin{split}
\vect{A} - \vect{B} = &\vect{e} (a_{t} - b_{t}) + \vect{i} (a_{x} - b_{x})+\\ &\vect{j} (a_{y} - b_{y}) + \vect{k} (a_{z} - b_{z}).
\end{split}
\end{equation}

The \textit{conjugate} of a fourvector changes the signs of the vector part:\\
\begin{equation}
\begin{split}
\overline{\vect{A}} 
 = \textbf{e} a_{t} - \textbf{i} a_{x} - \textbf{j} a_{y}- \textbf{k} a_{z}
\end{split}
\end{equation}

From this definition it is obvious that the result of summing a fourvector with its conjugate is a fourvector with only the scalar component different from zero. Divided by two, isolates the scalar component of a fourvector and serves to define the operator named the \textit{anticommutator} (similar to the scalar Hamilton's operator \textit{S}): $(\vect{A} + \overline{\vect{A}})/2=S\vect{A}$. Similarly, the result of subtracting the conjugate of a fourvector from itself is a \textit{pure fourvector} (that is, one whose scalar component is equal to zero). Divided by two serves to define the \textit{commutator} (similar to the vector Hamilton's operator \textit{V}): $(\vect{A} - \overline{\vect{A}})/2=V\vect{A}$\\

The \textit{complex conjugate} or hermitian conjugate of a fourvector changes the signs of the imaginary parts. Given the complex fourvector:
\begin{equation}
\begin{split}
\vect{A} = &\textbf{e} (a_{t} + i b_{t})+ \textbf{i} (a_{x} + i b_{x})+\\ &\textbf{j} (a_{y} + i b_{y})+ \textbf{k} (a_{z} + i b_{y})\\
&\tab \tab (10.15)\\
\end{split}
\end{equation}

Then its complex conjugate is:
\begin{equation}
\begin{split}
\vect{A}^{*} = &\vect{e} (a_{t} - i b_{t})+ \vect{i} (a_{x} - i b_{x})+\\ &\vect{j} (a_{y} - i b_{y})+ \vect{k} (a_{z} - i b_{y})
\end{split}
\end{equation}

Using relations \eqref{rel1} and \eqref{rel2}, the \textit{ fourvector product} is given by:
\begin{equation}
\begin{split}
\label{prod}
\vect{A}**\,\vect{B} = &\vect{e} (a_{t} b_{t} + a_{x} b_{x} + a_{y} b_{y} + a_{z} b_{z}) +\\
&\vect{i}\, (a_{t} b_{x} - a_{x} b_{t} + a_{y} b_{z} - a_{z} b_{y}) +\\
&\vect{j}\, (a_{t} b_{y} - a_{x} b_{z} - a_{y} b_{t} + a_{z} b_{x}) +\\
&\vect{k} (a_{t} b_{z} + a_{x} b_{y} - a_{y} b_{x} - a_{z} b_{t}).
\end{split}
\end{equation}

Using the notation of three-dimensional vector analysis we obtain a shorthand for the product. Regarding \vect{i}, \vect{j}, \vect{k} as unit vectors in a Cartesian coordinate system, we interpret the fourvector \vect{A} as comprising the scalar \textit{a} and the vector part \vect{a} = \vect{i} a$_{x}$ + \vect{j} a$_{y}$ + \vect{k} a$_{z}$. Then we write it in the simplified form \vect{A} = (\textit{a}, \vect{a}). With this notation, the product \eqref{prod} is expressed in the compact form:
\begin{equation}
\label{prodc}
\vect{A}**\,\vect{B} = (\textit{a b} + \vect{a}\cdot \vect{b}, \textit{a}\, \vect{b} - \vect{a}\, \textit{b} 
+ \vect{a} \times \vect{b})
\end{equation}

The following properties for the product are easily established:\\
\begin{enumerate}
\item If the scalar terms of both argument fourvectors of the product are zero then the resulting fourvector contains the classical scalar and vector products in its respective components.\\
\item The product is non-commutative. So, in general, there exist \vect{P} and \vect{Q} such that 
\vect{P}**\vect{Q} $\neq $ \vect{Q}**\vect{P}.

\item Fourvector multiplication is non-associative so, in general, 
\vect{P}**(\vect{Q}**\vect{R}) $\neq $ (\vect{P}**\vect{Q})**\vect{R} \\
Note that this is different from the Hamilton quaternions and the so-called Clifford Algebras, see for example \cite{aragon1997}.

\item The product of a fourvector by itself produces a result different from zero only in the first or ``scalar'' component, which is identified as the norm of the fourvector. In this sense this constitutes the classical dot product of vector calculus:
\begin{equation}
\vect{A}**\vect{A} = (a_{t}^{2} + a_{x}^{2} + a_{y}^{2} + a_{z}^{2}, 0, 0, 0)
\end{equation}

Note that this expression is substantially different with respect to the Hamilton quaternions, in which the square of a quaternion is given by
\begin{equation}
\vect{A}^{2} = (a_{t}^{2} - \vect{v}\cdot \vect{v},\; 2\, a_{t} \vect{v}),
\end{equation}

where \vect{v} represents the three vector terms of the quaternion.
Not only the scalar component has terms with the sign changed, but appears a non-zero term in the vector part of the quaternion. This has been a source of difficulty to apply Hamilton quaternions in Physics, which is overcome by the fourvectors.

\item The multiplicative inverse of a fourvector is simply the same fourvector divided by its norm.

\item Properties of the product and conjugates:
\begin{equation}
\overline{\vect{P}**\,\vect{Q}}=\vect{Q}**\,\vect{P}
\end{equation}
\begin{equation}
\overline{\vect{P}**\,(\vect{Q}**\,\vect{R})}=\vect{R}**\,(\overline{\vect{Q}}**\,\vect{P})
\end{equation}
\begin{equation}
\begin{split}
\vect{P}**(\overline{\vect{P}} **\vect{Q})
&=\overline{\vect{P}} **(\vect{P}**\vect{Q})\\
&=|\vect{P}|*\vect{Q}
\end{split}
\end{equation}
\begin{equation}
\begin{split}
(\overline{\vect{P}} **\vect{Q})**\vect{P}
&=(\vect{P}**\vect{Q})**\overline{\vect{P}}\\ 
&=|\vect{P}|*\overline{\vect{Q}} 
\end{split}
\end{equation}

With an operator notation: The product of two fourvectors is equal to the conjugate of the same product in reverse order:
\tab \vect{A}**\vect{B} = Conjugate[\vect{B}**\vect{A}]\\

\item For the case that ``\vect{r}'' is a rotor (a fourvector with $|\vect{r}| = 1$) then:\\
\tab \tab 

$\vect{r}**(\overline{\vect{r}} **\vect{Q})=\vect{Q}=\overline{\vect{r}} **(\vect{r}**\vect{Q})$
\tab \tab 

$(\overline{\vect{r}} **\vect{Q})**\vect{r}=\overline{\vect{Q}} =(\vect{r}**\vect{Q})**\overline{\vect{r}} $
\tab \tab 

$((((\vect{Q}**\vect{r})**\vect{r})**\overline{\vect{r}} )**\overline{\vect{r}} )=\vect{Q}$

(otherwise, if $|\vect{r}|$ is not equal to 1, the products of this numeral are equal to \vect{Q} or $\overline{\vect{Q}}$ multiplied by $|\vect{r}|$.)

\item The fourvectors do not contain the complex numbers, as is usually demonstrated for the Hamilton quaternions. The product of the fourvectors: (a, b, 0, 0) and (c, d, 0, 0) is\\ (ac+bd, ad-bc, 0, 0); also, the product of the fourvectors (a, \textit{i} b, 0, 0) and (c, \textit{i} d, 0, 0) is (ac-bd, \textit{i} ad-bc, 0, 0), whereas the product as complex numbers should be: (ac-bd, \textit{i} ad+bc, 0, 0)\\

\item Given the fourvectors \vect{A} and \vect{B}, the commutator:\\
\begin{align}
[\vect{A},\vect{B}]&= \ensuremath{\frac12} (\vect{A}**\vect{B} - \vect{B}**\vect{A})\\
&= (0, \textit{a} \vect{b} - \vect{a} \textit{b} + \vect{a}\times \vect{b})
\end{align}

gives a fourvector with zero scalar and with the vector part 
equal to the vector part of the fourvector product \vect{A}**\vect{B}.\\

For the curious, this commutator satisfies the properties of antisymmetry and linearity. The Jacobi identity is satisfied only for pure fourvectors.

\item Given two fourvectors, \vect{A} and \vect{B}, the anticommutator:\\
\begin{align}
\texttt{<}A,B\texttt{>}&= \ensuremath{\frac12} (\vect{A}**\vect{B} + \vect{B}**\vect{A}) \\
&= (\textit{a b} + \vect{a} \cdot \vect{b}, 0, 0, 0)
\end{align}

gives a fourvector with the scalar equal to the scalar of the fourvector product \vect{A}**\vect{B} and with the vector part equal to zero.\\

\item Product is left distributive over sum:

\vect{a}**(\vect{b} + \vect{c}) = \vect{a}**\vect{b} + \vect{a}**\vect{c}

\item Product is right distributive over sum:

(\vect{a} + \vect{b})**\vect{c} = \vect{a}**\vect{c} + \vect{b}**\vect{c}

\item The product of three ``pure'' fourvectors (defined as those whose scalar component is zero) can be expressed with the following vector products:\\

$\vect{a}**(\vect{b}**\,\vect{c}) = $\\
$( \vect{a} \cdot (\vect{b} \times \vect{c}),\; \vect{a} \times (\vect{b} \times \vect{c)} - \vect{a} 
* (\vect{b} \cdot \vect{c) )}$ \\

Where ``$\cdot$'' and ``$\times$'' are the standard vector
dot and cross products and ``*'' represents the product of the scalar $(\vect{b} \cdot \vect{c})$ by the vector \vect{a}. The scalar component of the result can be recognized as the volume of the parallelepiped having edges \vect{a}, \vect{b} and \vect{c}. Consequently, if the three vectors \vect{a}, \vect{b} and \vect{c} are in the same plane (or parallel to the same plane) then the scalar component of the resulting fourvector product is zero.\\

\item The following identity is also satisfied:\\
\tab \tab (\vect{a}**\vect{b}) ** (\vect{a}**\vect{b}) = (\vect{a}**\vect{a}) ** (\vect{b}**\vect{b})
\end{enumerate}

\subsection{Product with matrices}

Given two fourvectors, \vect{p} and \vect{q}:

\tab \vect{p}=(p0,p1,p2,p3),

\tab \vect{q}=(q0,q1,q2,q3),

their product can be obtained as

\vect{R}=\vect{p}**\vect{q}\\
The same product can be obtained multiplying the following matrix \vect{P} by the (four)vector \vect{q}:

\tab

$\vect{R} =\left[ 
\begin{array}{cccc}
\; p_{0}   & \; p_{1}   &\;  p_{2}  & \; p_{3}  \\
-p_{1}  & \; p_{0}   & -p_{3} & \; p_{2} \\
-p_{2}  & \; p_{3}   &\;  p_{0}  & -p_{1} \\
-p_{3} & -p_{2} & \; p_{1}  & \; p_{0}
\end{array}
\right] \left[ 
\begin{array}{c}
q_{0}  \\
q_{1}  \\
q_{2}  \\
q_{3}
\end{array}
\right] $\\

The \vect{P} matrix has the property\\
$\vect{P} \cdot \vect{P}^{T}=\vect{P}^{T}\cdot \vect{P}=\textbf{I}$, where $^T$ represents the transpose and \textbf{I} is the identity matrix. (More precisely the diagonal elements have the form: p$_{0}$$^{2}$+p$_{1}$$^{2}$+p$_{2}$$^{2}$+p$_{3}$$^{2}$, 
which are equal to 1 only if \vect{p} is a unit fourvector; else, in the diagonal is obtained the norm of the \vect{p} fourvector).

\subsection{The norm}

The \textit{norm} of a fourvector is defined by
\begin{equation}
|(a_{t},a_{x},a_{y},a_{z})|=a_{t}^{2}+ a_{x}^{2}+a_{y}^{2}+a_{z}^{2}
\end{equation}
It can be computed as the scalar component of the product of the fourvector by itself.\\
The norm satisfies the properties

\begin{equation}
|\overline{\vect{A}}|=|\vect{A}|
\end{equation}

\begin{equation}
|\vect{P}**\vect{Q}|=|\vect{Q}**\vect{P}|=|\vect{P}|*|\vect{Q}|
\end{equation}

The last property allows to conclude the following form of the \textit{four-squares theorem}:

\begin{equation}
\begin{split}
&(a_{0}^2+a_{1}^2+a_{2}^2+a_{3}^2)(b_{0}^2+b_{1}^2+b_{2}^2+b_{3}^2)=\\
&(a_{0}\, b_{0}+a_{1}\,b_{1}+a_{2}\,b_{2}+a_{3}\,b_{3})^{2}+\\
&(a_{0}\, b_{1}-a_{1}\,b_{0}+a_{2}\,b_{3}-a_{3}\,b_{2})^{2}+\\
&(a_{0}\, b_{2}-a_{1}\,b_{3}-a_{2}\,b_{0}+a_{3}\,b_{1})^{2}+\\
&(a_{0}\, b_{3}+a_{1}\,b_{2}-a_{2}\,b_{1}-a_{3}\,b_{0})^{2}
\end{split}
\end{equation}

\subsection{Identity fourvector}

The \textit{identity fourvector}, let us denote with \textbf{1}, has the scalar part equal to 1 and the vector part equal to zero: 
\textbf{1} = (1, 0, 0, 0). \\
It has the following properties, where ``\vect{p}'' is any fourvector: 

\tab \textbf{1}**\vect{p} = \vect{p}\\

\tab \vect{p}**\textbf{1} = $\overline{\vect{p}}$\\
As you can see, this is the left identity. It is possible to find the right identity of a fourvector but it is a little more complex. See the section ``Right factor of a fourvector''.

\subsection{Multiplicative inverse}

The \textit{multiplicative inverse} or simply \textit{inverse} of a fourvector \vect{P} is denoted by \vect{P}$^{-1}$.\\

The inverse of a fourvector \vect{P} is the same fourvector divided by its norm:
\begin{equation}
\vect{P}^{-1}=\vect{P}/|\vect{P}|
\end{equation}

The inverse operation satisfies the properties:

\tab \vect{P}**\vect{P}$^{-1}$=\textbf{1}
 
\tab \vect{P}$^{-1}$**\vect{P}=\textbf{1}

\tab (\vect{P}$^{-1}$)$^{-1}$ = \vect{P}

\tab (\vect{P} ** \vect{Q})$^{-1}$ = \vect{P}$^{-1}$ ** \vect{Q}$^{-1}$

\tab $\overline{\vect{P}^{-1}}=(\overline{\vect{P}})^{-1}$

Commutativity of products including inverses:\\
\noindent
\vect{P} ** (\vect{P}$^{-1}$ ** \vect{Q}) = \vect{P}$^{-1}$ ** (\vect{P} ** \vect{Q})\\
\noindent
(\vect{P}$^{-1}$ ** \vect{Q}) ** \vect{P} = (\vect{P} ** \vect{Q}) ** \vect{P}$^{-1}$ \\
\noindent
\vect{Q} = $\overline{\vect{P}}$**(\vect{P}$^{-1}$ ** \vect{Q}) = $\overline{\vect{P}^{-1}}$**(\vect{P} ** \vect{Q})\\
\noindent
$\overline{\vect{Q}}$ = (\vect{P}$^{-1}$ ** \vect{Q})**$\overline{\vect{P}}$ = (\vect{P} ** \vect{Q})**$\overline{\vect{P}^{-1}}$\\

\subsection{Scalar multiplication}

\textit{Scalar multiplication} – If \textit{c} is a scalar, or a scalar fourvector, and \vect{q}=(\textit{a}, \vect{v}) a fourvector, then
$c\, \vect{q} = (c,\vect{0})**\,\vect{q}$
$= (c,\vect{0})**(\textit{a}, \vect{v})
= (\textit{c} \textit{a} +\vect{0} \cdot \vect{v}, \textit{c} \vect{v} - \textit{a}\vect{0} +\vect{0} \times \vect{v})$\\
Simplifying:\\
$c\,(\textit{a}, \vect{v})= (\textit{c}\textit{a}, \textit{c}\, \vect{v})$

\subsection{Unit fourvector}

A \textit{unit fourvector} has the norm equal to 1. It is obtained dividing the original fourvector by its magnitude or absolute value, that is the square root of the norm. The product of two unit fourvectors is a unit fourvector. A unit fourvector can be represented with the use of trigonometric functions\\

\vect{\^{w}} = (\ensuremath{\pm}\;$\cos( \alpha )$ \ensuremath{\pm} \vect{\^{u}} $\sin ( \alpha )$) \\

where \vect{\^{u}} is in general a 3D vector of unit length.\\

The \textit{product of two unit fourvectors}:\\
Assume the unit fourvectors \vect{a} and \vect{b}:

\noindent
\vect{a} = ($\cos(\alpha)$, $\sin(\alpha)$, 0, 0)\\
\vect{b} = ($\cos(\beta)$, $\sin(\beta)$, 0, 0)\\

Its product is

\vect{a}**\vect{b} = ($\cos(\beta-\alpha)$, $\sin(\beta-\alpha)$, 0, 0)

so, if \vect{a} = \vect{b}, then the resulting fourvector is the \textit{identity fourvector}.\\

The \textit{inverse} of a unit fourvector is the same unit fourvector. This is because the product of the fourvector by itself produces the identity fourvector, or the norm, equal to 1, in the scalar component.

\subsection{Fourvector division}

The \textit{fourvector division} is performed by multiplying the ``numerator fourvector'', \vect{P}, by the ``denominator fourvector'', \vect{Q}, divided by its norm (or rather multiplying \vect{P} by the inverse of \vect{Q}):\\

$\vect{P}**\vect{Q}/|\vect{Q}| = \vect{P}**\vect{Q}^{-1}$\\

If \vect{P} and \vect{Q} are parallel ``vectors'' (pure fourvectors or with scalar part equal to zero), then the division produces, in the scalar part, the proportion between both vectors. For example:\\
\vect{P}=(1, 2, 3, 4)\\
\vect{Q}=(3, 6, 9, 12)

$\vect{P}**\vect{Q}/|\vect{Q}| = (1/3, 0, 0, 0)$

\subsection{Right factor of a fourvector}

Let us try to solve the following equation for ``\textbf{X}'' 

\tab \textbf{A} == \textbf{B} ** \textbf{X}

Let us assume that \textbf{A} and \textbf{B} are constant fourvectors and \textbf{X} is an unknown fourvector\\

\tab \textbf{A}=(a0,a1,a2,a3)

\tab \textbf{B}=(b0,b1,b2,b3)

\tab \textbf{X}=(x0,x1,x2,x3)

Where the components can be integer, rational, real or complex.

The solution for \textbf{X} can be obtained with the expression
\begin{equation}
\label{right}
\tab \textbf{X}=\overline{\textbf{B}}\,^{-1} ** \textbf{A}
\end{equation}
For example, let us try to determine what the value of \textbf{X} should be in order to satisfy the equation \textbf{A} == \textbf{B} ** \textbf{X}, when\\

\textbf{A}=(7,\,1,\,-3,\,5)\\

\textbf{B}=(1,\,3+\textit{i} 5,\,2,\,-1)\\

According to equation \eqref{right} the solution is \\

$\textbf{X}= (\frac{-3}{10}-\textit{i}\frac{2}{5}, \frac{9}{10}-\textit{i}\frac{4}{5}, \frac{12}{25}-\textit{i} \frac{53}{50}, \frac{9}{25}-\textit{i} \frac{21}{50})$

Replacing this solution into the product \textbf{B} ** \textbf{X} it can be verified that it reproduces the \textbf{A} fourvector.\\

\subsection{Left factor of a fourvector}

In a similar form, let us try to solve the following equation, where the unknown ``\vect{Y}'' is now a left factor of the constant fourvector \vect{B}:\\

\vect{A} == \vect{Y} **\, \vect{B}\\

\vect{Y} is obtained with the expression
\begin{equation}
\label{left}
\vect{Y}= \textrm{Hprod}[\overline{\vect{A}}, \vect{B}^{-1}]
\end{equation}

Where ``Hprod'' represents the product for the classical Hamilton quaternions or Grassman product.

For example, for the same fourvectors \vect{A} and \vect{B} from the previous example, let us apply the equation \eqref{left} and find\\

\noindent
$\vect{Y} = (\frac{4}{25}-\textit{i}\frac{1}{50},\,\frac{39}{50}-\textit{i}\,\frac{29}{25},-\frac{19}{25}+\textit{i}\frac{11}{50},-\frac{11}{50}+\textit{i}\frac{21}{25})$\\

Replacing this solution into the product\\
\textbf{Y} ** \textbf{B} it can be verified that it reproduces the \textbf{A} fourvector.

Note that the left and right factors of some fourvector such 
as \textbf{B} are different, although both factors have the same norm and satisfy the equality:\\

$\vect{X} **\, \vect{Y}^{-1} == \vect{X}^{-1} **\, \vect{Y}$\\

Both products, \vect{B} **\, \vect{X} and \vect{Y} **\, \vect{B} are equal to \vect{A}.

Formula \eqref{left} can be used to determine the single rotation fourvector that produces the same effect as a rotor. However, the results obtained are, in general, more complex than a classical rotor. Nevertheless, if we need the fourvector \vect{L}, which produces the same rotation of the fourvector \vect{p} as the rotor fourvector \vect{r}, that is:\\

$\vect{L} **\, \vect{p} = Rotate[\vect{p}, \vect{r}] = \vect{r} **\, (\overline{\vect{p}} **\, \vect{r})$\\

applying equation \eqref{left} we solve for \vect{L}:\\

$\vect{L} = \textrm{Hprod}[(\overline{\vect{p}} **\, \vect{r}) **\, \vect{r}], \vect{p}^{-1}]$\\

Where ``Hprod'' is the Grassman product.\\

\subsection{Solution of quadratic fourvector polynomials}

There can be an infinite number of solutions for a quadratic 
fourvector polynomial. Consider the quadratic equation \\

$\vect{q}^{2} ==\vect{1}$. \\

Then, the fourvector \vect{q} = (cos(x), sin(x),0,0), where x is any real number is a collection of solutions for this equation 
because the norm of \vect{q} is 1:\\

\vect{q} ** \vect{q} == \vect{1}\\

So the above choice for \vect{q} satisfies the equation $\vect{q}^{2} ==\vect{1}$ for all real values of x.\\

When there is a solution of a quadratic equation, it can be computed as in the following.\\
Assume a quadratic polynomial of the form:\\

\vect{q}**\vect{q}+\vect{q}**\vect{j}==\vect{k}\\

where:\\

\vect{q}=(q0,q1,q2,q3) is an unknown fourvector and\\

\vect{j}=(0,-1,1,0) and \vect{k}=(-1,0,0,1) are constant fourvectors

From here, the four equations, obtained equating the four components, are:\\

\noindent
1+q0$^{2}$- q1+ q1$^{2}$+ q2+ q2$^{2}$+ q3$^{2}$==0,\\
-q0+q3==0,\\
q0+q3==0,\\
-1-q1-q2==0.\\

This system of equations has two solutions for the four components of the fourvector:\\

\begin{align}
\vect{q}_{1}&=(0, -1-\textit{i}/\sqrt{2},\textit{i}/\sqrt{2},0)\\
\vect{q}_{2}&=(0, 1+\textit{i}/\sqrt{2}, -\textit{i}/\sqrt{2},0)
\end{align}\\

Replacing $\vect{q}_{1}$ (or $\vect{q}_{2}$) by its value, in the following expression, which is the left hand side of the given equation, \vect{q}**\vect{q}+\vect{q}**\vect{j}, the value returned is: (-1,0,0,1),\\

Which is the value of the right hand side.\\

For a comparable quadratic equation, but now affecting the ``\vect{j}'' fourvector to the left of ``\vect{q}'' instead of to the right,\\

\vect{q}**\vect{q} + \vect{j}**\vect{q}==\vect{k}\\

The two solutions are:
\begin{align}
\vect{q}_{1}&=(0, -\textit{i}/\sqrt{2},1/2(-2+\textit{i}\sqrt{2}),0)\\
\vect{q}_{2}&=(0, \textit{i}/\sqrt{2}, 1/2(-2-\textit{i}\sqrt{2}),0)
\end{align}

Which, replacing in \tab \vect{q} ** \vect{q} + \vect{j} ** \vect{q}

produce the value:\tab (-1,0,0,1)\\

Note that the fourvectors form a \textit{division algebra} since they have a (left) multiplicative identity element \vect{1}$\neq$\vect{0} and every non-zero element \vect{a} has a multiplicative inverse (i.e. an element \vect{x} with \vect{a\,x} = \vect{x\,a} = \vect{1}).\\

\section{Fourvector rotation}
\label{qrotation}
Mathematically, a rotation is a linear transformation that leaves the norm invariant. These are called orthogonal transformations.\\

There are several methods to represent rotation, besides quaternions, including Euler angles, orthonormal matrices, Pauli spin matrices, Cayley-Klein parameters, and extended Rodrigues parameters. \\

From the several ways to represent the attitude of a rigid body one of the most popular is a set of three Euler angles. Some sets of Euler angles are so widely used that they have names, such as the \textit{roll}, \textit{pitch}, and \textit{yaw} of an airplane. The main disadvantages of Euler angles are that certain functions of Euler angles have ambiguity or singularity for certain angles. This produces, for example, the so-called ``gimbal lock''. Also, they are less accurate than unit quaternions when used to integrate incremental changes in attitude over time \cite{diebel03}.\\

The handling of rotations by means of quaternions has constituted the technical foundation of modern inertial guidance systems in the aerospace industry for the orientation or ``attitude'' of satellites and aircrafts. This task is to be shown in the following that should be carried out by fourvectors.

Many graphics applications that need to carry out or interpolate the rotations of objects in computer animations have also used quaternions because they avoid the difficulties incurred when Euler angles are used. The form to replace by fourvectors is not performed in the present paper, although it should be perfectly possible.\\

The use of matrices is neither intuitive for the localization of the axis of rotation nor efficient for computation. But one of the most important disadvantages is the associativity of both the matrix and Hamilton's quaternion products where, for example, $\vect{A}\cdot(\vect{B}\cdot\vect{C})$ is equal to  $(\vect{A}\cdot\vect{B})\cdot\vect{C}$. In fact it is rather well known that the composition of rotations, when either matrices or Hamilton quaternions are used, is associative. This means that these mathematical tools produce the same result no matter what the grouping of a sequence of two or more rotations. This poses a serious technical problem to the engineers who need to distinguish between two sequences of the same rotations. To illustrate with an example, let us assume that you are piloting an airplane with a local frame of reference whose origin is attached at the center of the airplane. Assume that, at a certain instant, the ``x'' axis, which is pointing toward the front of the plane, is horizontal according to an observer in the earth, let us say directed toward the North pole, the ``y'' axis points to the right wing, that is pointing to the East, and the ``z'' axis points toward the earth's center. If under these conditions you maneuver to produce a $90^\circ$ roll (a quarter circle clockwise rotation about the local x axis) followed by a $90^\circ$ yaw (rotation ``to the right'' around the local z axis) your plane should be falling perpendicularly to the earth; but if you interchange the order of these rotations, that is first the yaw and then the roll, this should put your plane in a horizontal heading toward the East, although with the right wing pointing toward the earth's center.\\

To rotate a vector \vect{u} around an Euler vector \vect{n} through an angle $\theta$, one has to apply the following equation (see Fig. 1 and references \cite{salamin1979}, \cite{stevens03}): 
\begin{equation}
\vect{v} = \vect{u} cos\theta + \vect{n}(\vect{n} \cdot \vect{u}) (1 - cos\theta) + (\vect{n} \times \vect{u}) sin\theta
\end{equation}

\begin{figure}[hbtp]
\begin{center}
\includegraphics[width=3.1in]{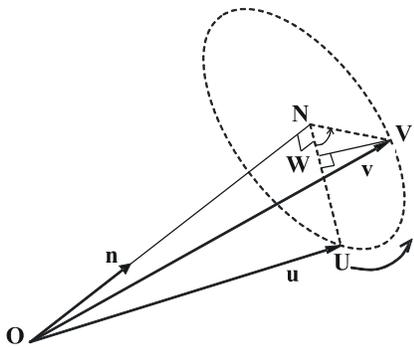}
\caption{Graphic of a rotation}
\end{center}
\end{figure}

According to Silberstein, \cite{silberstein1912}: ``It has been remarked by Cayley, as early as in 1854, that the rotations in a four-dimensional space may be effected by means of a pair of quaternions applied, one as a prefactor and the other as a postfactor, to the quaternion \vect{u} whose components are the four coordinates of a space-point, say $\vect{v} = \vect{a}\,\vect{u}\, \vect{b}$''. This phrase applies directly to fourvectors if ``quaternion(s)'' is replaced by ``fourvector(s)''. \\
In the case of pure rotation, \vect{a} and \vect{b } must be either unit-fourvectors or the norm of their product must be 1: 
			$|\vect{a}|* |\vect{b}| =1$.

It follows, from the rule: 
\begin{equation}
|\vect{a}**\,\vect{b}| = |\vect{a}| * |\vect{b}|, 
\end{equation}
the multiplication of the fourvector being rotated by unitary fourvectors \vect{a} and \vect{b}, effects an orthogonal transformation.\\
This form can be simplified so instead of two different unitary fourvectors is selected only one, let us name it \vect{r}.

A possible fourvector \vect{r} that produces the rotation of any fourvector \vect{V} about a certain axis ``\vect{n}'', through an angle $\theta$, has the form (\cite{salamin1979}, \cite{dam98}, \cite{goddard97} \cite{hart1994}, \cite{hoffmann05}, \cite{fontijne00}):
\begin{align}
&\vect{r} = (cos(\frac{\theta}{2}), n_{x}\;sin(\frac{\theta}{2}), n_{y}\;sin(\frac{\theta}{2}), n_{z}\;sin(\frac{\theta}{2}))\\
\nonumber
&\textrm{or simply}\\
&\vect{r} = (cos(\frac{\theta}{2}), \vect{n}\; sin(\frac{\theta}{2}))
\end{align}
The rotation is carried out with the following product:
\begin{equation}
\vect{V'}=\vect{r}**\,(\overline{\vect{V}}**\,\vect{r})
\end{equation}
The rotation operand needs to be conjugated. This is different with respect to the rotation using Hamilton quaternions \cite{silberstein1912}, where the inverse or the conjugate of the second rotor is needed.\\
Fig. 2 shows the vector \vect{V} rotated around the unit vector \vect{k}, through an angle $\theta$, with rotor \vect{r}.
\begin{figure}[hbtp]
\begin{center}
\includegraphics[width=3.1in]{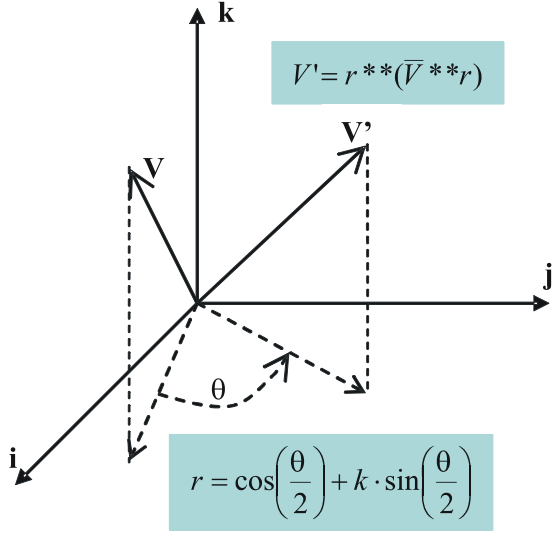}
\caption{Example rotation}
\end{center}
\end{figure}
The rotation operator is linear. It can be proved that the rotation of either the product or the sum of two fourvectors \vect{s} and \vect{t} with the rotor \vect{r}, is the same as, respectively, the product or the sum of the rotations of each fourvector:
\begin{align}
&Rotate[\vect{s}**\,\vect{t},\vect{r}] \equiv Rotate[\vect{s},\vect{r}]**\,Rotate[\vect{t},\vect{r}]\\
&Rotate[\vect{s}+\,\vect{t},\vect{r}] \equiv Rotate[\vect{s},\vect{r}]+\,Rotate[\vect{t},\vect{r}]
\end{align}
Let us define the following example rotation fourvectors, or rotors, whose norm is the unit: $\vect{q}_i**\,\vect{q}_i=\textbf{1}$, and cause rotations about the \textit{x} axis:

\begin{align}
\vect{q}_1 &= (\pm cos(\alpha/2), \pm sin(\alpha/2), 0, 0)\\
\vect{q}_2 &= (\pm cosh(\alpha/2), \pm i\, sinh(\alpha/2), 0, 0)\\
\vect{q}_3 &= (\sqrt{\frac{\gamma + 1}{2}}, \textit{i}\, \sqrt{\frac{\gamma - 1}{2}}, 0, 0) 
\end{align}
where $\gamma$ is the Lorentz contraction factor.\\

The rotation fourvector $\vect{q}_{3}$ was obtained by transforming the following fourvector

(\ensuremath{\pm} $\gamma$, \ensuremath{\pm} \textit{i} $\beta \; \gamma$, 0, 0)

in such a way that it produces a rotation of half the hyperbolic 
angle, as with $\vect{q}_{2}$.

Let us multiply any one of the previous fourvectors by the following one that represents a differential of interval:

\vect{ds} = (c dt, \textit{i} dx, \textit{i} dy, i dz)\\

The products are of the form:
\begin{equation}
Rotate[\vect{ds}, \vect{q}_{i}]=\vect{q}_{i} **\, (\overline{\vect{ds}} **\, \vect{q}_{i})
\end{equation}

where \vect{q$_{i}$} is anyone of the above list of unit fourvectors.\\
Then, the norm of the result is the square of the differential 
of interval: 

ds$^{2}$ = c$^{2}$ dt$^{2}$ - dx$^{2}$ - dy$^{2}$ - dz$^{2}$ 

This means that any one of these transformations (rotations) preserves the interval invariant.

But first the rotation of a real fourvector\\ $\vect{a}=(a_{t},a_{x},a_{y},a_{z})$ with the rotor $\vect{q}_{1}$ produces the classical formulas for rotation of a vector about the \textit{x} axis:
\begin{equation}
\begin{split}
Rotate[&\vect{a},\vect{q}_{1}]=(a_{t},\,a_{x},\\
& a_{y} \cos(\alpha) - a_{z} \sin(\alpha),\;\\  
&a_{z} \cos(\alpha) + a_{y} \sin(\alpha))
\end{split}
\end{equation}

The opposite rotation can be done using as rotor the inverse of $\vect{q}_{1}$, which changes the sign of the angle $\alpha$. For example, if we rotate the last result with the conjugate or the inverse of $\vect{q}_{1}$ then the original fourvector \vect{a} is recovered.\\

The rotation of the differential of interval with $\vect{q}_{2}$ gives the complex fourvector:
\begin{equation}
\begin{split}
Rotate[&\vect{ds},\vect{q}_{2}]=(c\, dt, \; \textit{i}\, dx,  \; \\
&dz \sinh(\alpha)+\textit{i}\, dy \cosh(\alpha), \\
-&dy \sinh(\alpha) +  \; \textit{i}\, dz \cosh(\alpha))
\end{split}
\end{equation}

The rotations with $\vect{q}_{3}$ produce Lorentz boosts. Let us apply to \vect{ds}: \\
\begin{equation}
\begin{split}
Rotate[&\vect{ds},\vect{q}_{3}]=(c dt , \; \textit{i}\; dx, \; \\
&\textit{i}\; \gamma  \;(dy- \textit{i}\,  \beta \, dz), \;\\ &\textit{i}\; \gamma \;(dz+ \textit{i}\, \beta \,dy))
\end{split}
\end{equation}

If the fourvector to be rotated is the previous \vect{a} and the rotor fourvector is\\

$\vect{r} = (\cos(\alpha /2), 0, 0, \sin(\alpha/2) )$\\

Then the double rotation: \\

$\vect{r}**(\vect{r}**(\vect{a}**\vect{r})**\vect{r})$\\

Is equal to a single rotation through the double angle $2\alpha $. The result is:\\
\begin{equation}
\begin{split}
(a_t,\,&a_x \cos(2 \alpha)-\,a_y \sin(2 \alpha),\\
&a_y \cos(2 \alpha)+\,a_x \sin(2 \alpha),\,a_z)
\end{split}
\end{equation}

\subsection{Composition of rotations}

The rotation through an angle $\alpha $ followed by another rotation through an angle $\beta $ is equivalent to a single rotation through an angle $\alpha +\beta $:\\

Assume, for example that the rotations are produced by application of the following rotors:
\begin{align}
roth1&=(cosh(\alpha/2),i \,sinh(\alpha/2),0,0)\\
roth2&=(cosh(\beta/2),i \,sinh(\beta/2),0,0)
\end{align}

Let us apply these rotors over the fourvector \vect{M}=(a,b,c,d) with the operation:
\begin{equation}
\vect{M1}=Rotate[\vect{M},roth1]
\end{equation}
followed by the following rotation:
\begin{equation}
\vect{M2}=Rotate[\vect{M1},roth2]
\end{equation}

We obtain the following result:
\begin{equation}
\begin{split}
(a, i\,b, &i\,c\, cosh(\alpha +\beta) + d\, sinh(\alpha +\beta),\\
&i\,d\, cosh(\alpha +\beta) - c\, sinh(\alpha +\beta))
\end{split}
\end{equation}

Which is identical to the rotation produced by directly applying over \vect{M} the rotor:
\begin{equation}
roth=\big{(}cosh(\frac{\alpha +\beta}{2}),i\,sinh(\frac{\alpha +\beta}{2}),0,0\big{)}
\end{equation}

\subsection{Reflections}

Let us show how the reflection in a plane with unit normal \textbf{a} can be done (see next figure)\\

\begin{figure}[htbp]
\begin{center}
\includegraphics[width=3.1in]{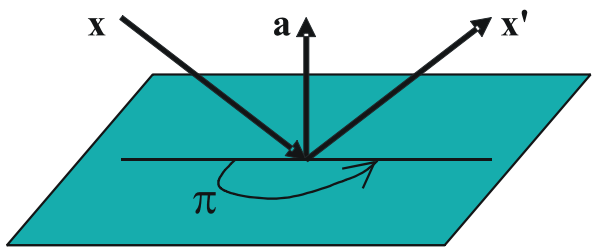}
\caption{Reflection}
\end{center}
\end{figure}

the normal vector \textbf{a}, which defines the direction of the plane, generates a reflection of the vector \textbf{x} if we rotate \textbf{x} around the vector \textbf{a} through an angle of $\pi $ radians and then change every sign of \textbf{x} (otherwise we end up with an arrow pointing at the same point as \textbf{x}). This rotation is done by fixing at $\pi $ radians the angle $\alpha $ of rotation of a rotor of the form \vect{q1} of section \ref{qrotation}, i.e.\\

q[Cos[$\alpha /2$],\textbf{a} Sin[$\alpha /2$]].\\

Consequently, the cosine term disappears and the sine term 
becomes equal to 1, with which we are left with the vector \textbf{a} alone, as rotor.

The vector \vect{x}, after reflection, is:

$\vect{x'} = - \vect{a}**(\overline{\vect{x}}**\vect{a)}$

To simplify this expression, let us note that the rotation through $\pi $ radians clockwise is the same as the rotation through $\pi $ radians counterclockwise, so we could include conjugations of 
the vector \vect{a}. Also, the conjugation of the vector \vect{x} can 
be canceled with the negative sign, so the final equation is:

\vect{x'} = \vect{a}**(\vect{x}**\vect{a)}.\\

As was suggested at the end of the previous section, the rotations 
can be composed. So that, by multiple application of this reflection 
formula it is possible to compute the vector \vect{x'} that describes the direction of emergence of a ray of light initially propagating with direction \vect{x} and reflecting off a sequence of plane surfaces with unit normals \vect{a}$_{\mathbf{1}}$; \vect{a}$_{\mathbf{2}}$; \dots  ;\vect{a}$_{\mathbf{n}}$:

\vect{x'} =\vect{a}$_{\mathbf{n}}$**(...(\vect{a}$_{\mathbf{2}}$**((\vect{a}$_{\mathbf{1}}$**(\vect{x**a}$_{\mathbf{1}}$))**\vect{a}$_{\mathbf{2}}$)...\vect{a}$_{\mathbf{n}}$)

\begin{figure}[htbp]
\begin{center}
\includegraphics[width=3.1in]{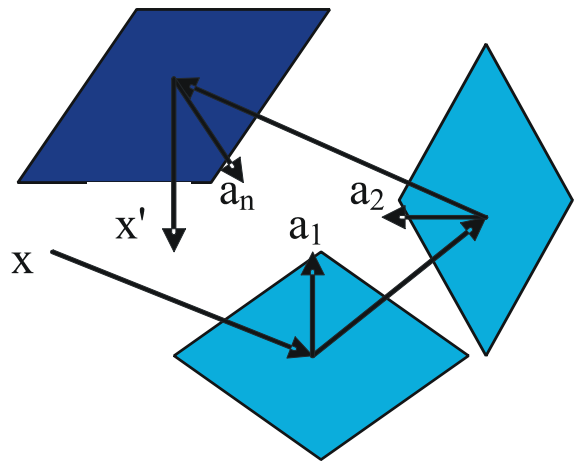}
\caption{Reflections}
\end{center}
\end{figure}

Hestenes [15] shows other applications for rotations and reflections, for example in crystals. Note that in his ``geometric algebra'' Hestenes needs the negative sign that we don't need.\\

\section{Conclusions}

The mathematical structure of quaternions has always been considered as more appropriate than the simple vectors to represent the real physical variables. Nevertheless, the quaternions were dismissed for the difficulties and complications produced by their quaternion product. With the new product, suggested in the present paper for fourvectors, all those difficulties disappear. Of course there must be a delicate balance between the correct mathematical tools and the real physical objects being studied and handled. One has to also be aware that mathematics clearly affects the ontology of physics \cite{gingras01}. \\

The fourvector algebra proposed in the present paper seems to be the correct mathematical tool to study the fundamental physical variables and their describing equations.\\

This new mathematical structure is an extension of the classical vectors. Its simplicity contributes to the possibility of more extended and fruitful use in all branches of science.\\

The applications of the Hamilton quaternions for rotations in three dimensions have been the more extended in current Physics. Such uses, as well as reflections, are still permitted by the fourvectors. The applications to Lorentz boosts had problems with the old quaternions, so this area opens up for the scientists.\\

\end{document}